\documentclass[twocolumn,prd,nofootinbib,aps,floats,floatfix,amsmath,amssymb,secnumarabic]{revtex4} %
\usepackage[dvips,final]{graphicx}
\usepackage{amsmath}
\usepackage{amsfonts}
\usepackage{amssymb}
\usepackage{latexsym}
\usepackage{graphicx}
\usepackage[english]{babel}
\usepackage{multirow}
\usepackage{float}
\usepackage{url}
\usepackage{hyperref}
\usepackage{slashed}

\def\sfrac#1#2{{\textstyle{#1\over #2}}}
\newcommand{\be}{\begin{equation}}
\newcommand{\ee}{\end{equation}}
\newcommand{\ba}{\begin{array}}
\newcommand{\ea}{\end{array}}
\newcommand{\bea}{\begin{eqnarray}}
\newcommand{\eea}{\end{eqnarray}}
\newcommand{\sss}{\scriptscriptstyle}

\def\roughly#1{\mathrel{\raise.3ex\hbox
{$#1$\kern-.75em\lower1ex\hbox{$\sim$}}}}
\def\lsim{\roughly<}
\def\gsim{\roughly>}

\newcommand{\de}{{\mathbf e} }
\newcommand{\pd}{{\mathbf p} }
\newcommand{\dH}{{\mathbf H} }

\begin{document}

\title{Millicharged Atomic Dark Matter}
\author{James M. Cline}
\author{Zuowei Liu}
\author{Wei Xue}
\affiliation{Department of Physics, McGill University,
3600 Rue University, Montr\'eal, Qu\'ebec, Canada H3A 2T8}

\begin{abstract}   We present a simplified version of the atomic dark matter scenario, in which
charged dark constituents are bound into atoms analogous to
hydrogen by a massless hidden sector U(1) gauge interaction. Previous
studies have assumed that interactions between the  dark sector and
the standard model are mediated by a second, massive $Z'$ gauge boson,
but here we consider the case where only a massless $\gamma'$
kinetically
mixes with the standard model hypercharge and thereby mediates direct
detection.  This is therefore the simplest atomic dark matter model
that has direct interactions with the standard model, arising from
 the small electric charge for the dark constituents
induced by the kinetic mixing.  We map out the parameter space that is consistent with
cosmological constraints and direct searches, assuming
that some unspecified mechanism creates the asymmetry that gives the
right abundance, since the dark matter cannot be a thermal relic in
this scenario.   In the special case where the dark ``electron'' and
``proton'' are degenerate in mass, inelastic hyperfine transitions can
explain the CoGeNT excess events.  In the more general case, elastic
transitions dominate, and can be close to current direct detection
limits over a wide range of masses.

\end{abstract}
\maketitle
\newpage
\section{Introduction}
In recent years there has been increased interest in dark matter
models in which the dark sector has some of the richness of the
visible sector, such as hidden gauge interactions 
\cite{Foot:2003iv}-\cite{Feldman:2007wj} and flavor.  A natural
possibility to consider is an unbroken U(1) gauge
symmetry that would give rise to bound states, {\it i.e.,} atomic
dark matter \cite{Goldberg:1986nk}-\cite{Kaplan:2011yj}.  In this case there should be a further resemblance to
the visible sector in that the dark matter must be asymmetric 
\cite{Kaplan:2009ag} in order
to have the right abundance; otherwise the U(1) coupling must be so
weak that recombination in the dark sector does not occur efficiently
and the would-be atoms remain predominantly ionized 
\cite{Ackerman:2008gi}.  

Atomic dark matter can have interesting properties with respect to
direct detection, because of the possibility of inelastic scattering
to excited states of the atom, notably through hyperfine transitions.
In previous studies it has been shown that inelastic transitions can
help to reconcile the DM interpretation of CoGeNT events 
\cite{Aalseth:2010vx} with null
results from Xenon10 \cite{Angle:2011th}.  In refs.\ \cite{Kaplan:2009de,Kaplan:2011yj}
it was assumed that the hyperfine transitions were mediated by the
kinetic mixing of the photon with a massive dark vector boson that
couples to the axial vector current of the DM.  In the present work we
explore a simpler possibility \cite{Goldberg:1986nk}: one can rely upon mixing of the
photon with the massless $\gamma'$ that is already present due to the
unbroken U(1) gauge symmetry.  This is a very economical model, while
still endowed with the rich phenomenology of the atomic DM scenario.

Because of the kinetic mixing, the constituents of the dark atoms
acquire small electric charges $\epsilon e$; thus the interactions
that give rise to direct detection are electromagnetic.  We show that
direct searches in fact give the strongest bounds on
$\epsilon$ in this model; thus such detections could be imminent.
In fact in the special case where the two constituents have equal
mass, we offer an interpretation for the excess events reported
by CoGeNT, relying upon inelastic interactions.

\section{Direct detection}
We follow the notation of ref.\ \cite{Kaplan:2009de} by denoting the 
dark analogues of the proton, electron, and hydrogen atom
by $\mathbf p$, $\mathbf e$, $\mathbf H$.  The Lagrangian is
\bea
	{\cal L} &=&\bar{\mathbf e}( i\slashed{D}'-m_{\mathbf e}) {\mathbf e}
	+\bar{\mathbf p}( i\slashed{D}'-m_{\mathbf p}) {\mathbf p}
	\nonumber\\
	&&-\sfrac14 F_{\mu\nu}F^{\mu\nu} -\sfrac14 \tilde F'_{\mu\nu}
	\tilde F'^{\mu\nu} +\sfrac12\tilde\epsilon F_{\mu\nu}{\tilde F'}_{\mu\nu}
\eea
where $F$ is the electromagnetic field strength and $\tilde F'$ is that
of the massless $\gamma'$.  $D'$ is the covariant derivative with respect
to $F'$: $D' = \partial \pm i g A'$, where $g$ is the U(1)$_d$ dark
coupling constant. 
We ignore the small mixing of the $\gamma'$
with the $Z$ boson, and hence refer to $F$ as the electromagnetic rather than the
hypercharge field strength.
$\tilde\epsilon$ is the gauge kinetic mixing
parameter.  The gauge boson kinetic terms can be diagonalized to first order
in $\tilde\epsilon$ by letting $\tilde{F'} = F' + \tilde\epsilon F$.  Then $A'_\mu$
couples only to the dark current $g J_d^\mu$ while the photon $A_\mu$
couples to $e J_{em}^\mu + \tilde\epsilon g J_d^\mu$ 
\cite{Holdom:1985ag,Burrage:2009yz}. 
The DM particles thus acquire
millicharges $\tilde\epsilon g \equiv \epsilon e$ 
under the electromagnetic U(1).  We will refer to
$\epsilon$ rather than $\tilde\epsilon$ in the remainder of the paper.

\subsection{$m_\de \ll m_\pd$ case}  

{\bf Dark atom interactions.} The low-energy
interactions of $\mathbf H$ are screened due to its net
charge neutrality.
Let us first consider the generic regime where $m_\de \ll m_\pd$.
In this limit, the Fourier transform of the $\mathbf H$ electric charge density 
is given by
$\tilde\rho_{\mathbf H} \cong \epsilon e\, {a'_0}^2\, q^2/2$
at low wave-number $q \ll 1/a'_0$, where $a'_0 \cong 1/(\alpha' m_\de)$
is the Bohr radius of $\dH$, and $\alpha' = g^2/4\pi$. We assume
that $m_{\de}\ll m_\pd$ in our approximation for $a'_0$.  The factor
of $q^2$ in $\tilde\rho$ cancels the factor of $1/q^2$ coming from 
the gauge boson propagator in Coulomb gauge, so that the scattering 
of $\dH$ on a proton in a DM detector will not be long-range, but will
instead appear as a contact interaction, with
\be
	{\sigma_{p}} = 
	4\pi\, \alpha^2\epsilon^2 \mu^2 {a'_0}^4
\label{sigmap}
\ee
where $\mu$ is the reduced mass of the $p$-$\dH$ system.  This
expression relies upon the Born approximation, whose validity depends
upon the properties of the central potential experienced by $p$
due to $\dH$:
\be
	V = {\epsilon\alpha\over a'_0}
	\, e^{-2r/a'_0}\left( 1 + {a'_0\over r}\right)
\ee
The Born approximation is justified if $|V(a'_0)|\ll 1/(\mu {a'}_0^2)$,
which implies
\be
	\epsilon {\alpha\over\alpha'} \ll {m_\de\over \mu}
	 =  {m_\de\, (m_p+ m_\dH) \over m_p\, m_\dH}
\label{born_cond}
\ee
We will see that this condition is satisfied for parameters of
interest for direct detection.
In comparing $\sigma_p$ to direct detection bounds, we must take into
account  that $\dH$
interacts only with protons and not all nucleons.  This weakens the
experimental limit on $\sigma_p$ by a factor of $(Z/A)^2$ for a target
with atomic number and weight $Z,A$.  For Xenon, 
$(Z/A)^2 = 0.17$.   We define $\sigma_{p,\rm eff} = (Z/A)^2 \sigma_p$ to
facilitate comparison with the Xenon excluded region.

\begin{figure}[ht]
\hspace{-0.4cm}
\includegraphics[width=0.5\textwidth]{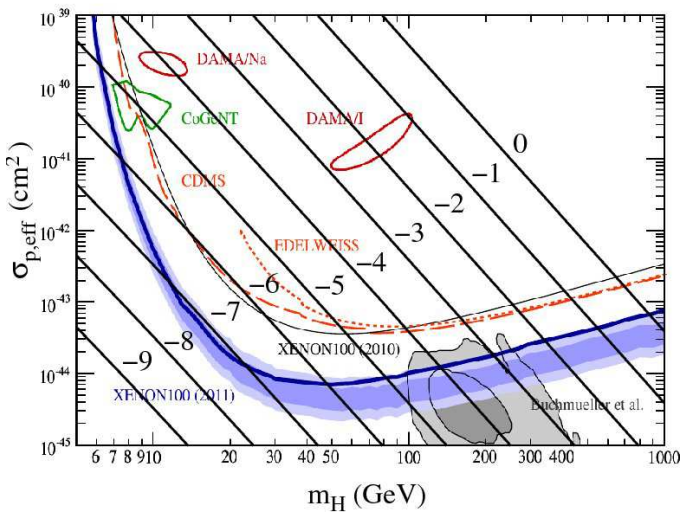}
  \caption{Diagonal lines: contours of constant $\beta$,
eq.\ (\ref{betaeq}), in the $m_\dH$-$\sigma_{p,\rm eff}$ plane.
Lines are labeled by the value of $\log_{10}\beta$.
Background shows limits from the Xenon100 experiment
\cite{Aprile:2011hi}.}
\label{xenon}
\end{figure}

\begin{figure}[ht]
\hspace{-0.4cm}
\includegraphics[width=0.5\textwidth]{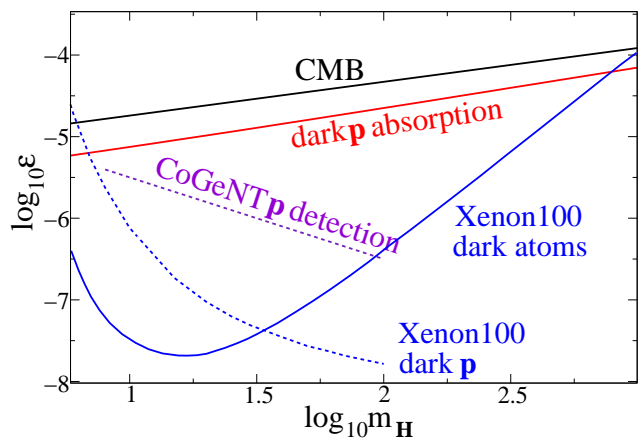}
  \caption{Lowest curve: values of $\epsilon$ that saturate the Xenon100 direct
detection bound, as a function of $m_\dH$, for elastic scattering
of atoms in the model with
$\alpha'=m_\de/m_\pd = 0.1$.  Upper curves: upper limits on 
$\epsilon$ from 
CMB (ref.\ \cite{McDermott:2010pa}), capability of $\pd$ ions to 
penetrate 1 km of rock (this work), CoGeNT limit (adapted from \cite{McDermott:2010pa})
and Xenon100 limit (our estimate) from
detection of $\pd$ ions.  Note that the ion
detection limits may not apply (see section 2.1).
\vspace{-0.5cm}}
\label{epsfig}
\end{figure}

The cross section for direct detection is proportional to the 
combination
\be
	\beta \equiv {\epsilon^2\over \alpha'^4}\,
		{(1+ x_\de)^4\over x_\de^4}\, 
\label{betaeq}
\ee
when we reexpress $\mu$ and $a'_0$ in
terms of $m_\de$ and $m_\dH$, where 
$x_{\de} \equiv m_\de/m_\pd \cong m_\de/m_\dH$.
The values of $\beta$ that are interesting for direct detection can
be read from fig.\ \ref{xenon}, where contours of constant $\beta$
(labeled by the value of $\log_{10}\beta$), are plotted in the 
$m_\dH$-$\sigma_{p,\rm eff}$ plane, on top of constraints from the
Xenon100 experiment.  To give more concrete examples, let us take
$\alpha' = 0.1$ and $x_\de = 0.1$, which tend to give a small
ionization fraction
 and therefore are more robust with respect to structure
formation constraints.  (The ionization fraction $f$ is computed
in analogy to that of visible hydrogen in ref.\ \cite{Kaplan:2009de},
and goes roughly as $f\sim 10^{-10}\alpha'^{-4}m_\de m_\pd$ GeV$^{-2}$.)
The values of $\epsilon$ needed to saturate
the Xenon100 bound \cite{Aprile:2011hi} for a range of $m_\dH$ are
shown in figure 
\ref{epsfig}. These scale as $x_\de^2\alpha'^{2}$ for different
values of $x_\de$ and $\alpha'$.  The values shown in fig.\ \ref{epsfig}
easily satisfy condition (\ref{born_cond}).
	
In principle, inelastic scattering can also occur, in which the
internal atomic state changes.  The lowest energy excitation available
is the hyperfine transition, with energy gap $\Delta E\cong \sfrac83
\alpha'^4 m^2_\de/m_\pd$.  For our fiducial parameters $\alpha'=x_\de=0.1$,
this would correspond to $\sim 30$ keV if $m_{\dH}\cong 10$ GeV. Although
this may be an interesting value for direct detection, 
 the rate of such
transitions is suppressed compared to the elastic ones by a factor of
$\alpha'^4 x_\de^2(m_\dH/m_p)^2$ 
 so they make a subdominant contribution if $m_\dH\lsim$ TeV.
Transitions that excite the $\de$ orbital state are less suppressed
by powers of $\alpha'$, but have a larger energy gap, and so are
also irrelevant.	

{\bf Dark ion interactions.} An interesting feature of atomic dark
matter is that a small fraction remains in the ionized state, which
has a larger cross section on nucleons than does the atomic state,
because its charge is unscreened.  Therefore one might question
whether the scattering of ionic DM could dominate the direct detection
signal, even though it is a subdominant component.  Ref.\
\cite{McDermott:2010pa} has argued that the ionized component would
necessarily have been blown out of the galactic disk by supernova
shock waves \cite{Chuzhoy:2008zy} if $\epsilon$ was in the range required for direct
detection.  The galactic magnetic field subsequently shields the disk
from being repopulated by the millicharged ions, unless
\cite{Foot:2010yz} the strong scatterings between the ions themselves
sufficiently  randomize their directions contrary to the Lorentz force
from the magnetic field.  Adapting the estimate (8) of
\cite{Foot:2010yz} for the distance scale $l$, over which
randomization occurs, to our models with $\alpha' = 0.1$, $m_\de/m_\pd
= 0.1$ and using the $m_\dH$-dependent ionization fraction from
fig.\ 1 of ref.\ \cite{Kaplan:2009de}, we find that $l \cong 300$ pc, which
is larger than the height of the galactic disk and thus ineffective for
overcoming magnetic shielding.

Nevertheless, in case there may be some other way of evading this
argument, we indicate by the dashed line of fig.\ \ref{epsfig} the 
value of $\epsilon$ at which dark ionic scattering would saturate the CoGeNT
signal, which is adapted from ref.\ \cite{McDermott:2010pa} by taking
into account the ionization fraction $f$ mentioned above.  Thus even if the ions
penetrate to the earth, their signal is much weaker than that of the
atoms unless $m_\dH > 100$ GeV.  For consistency one must also check
that dark ions can penetrate $\sim 1$ km of rock, which we have done
along the lines of refs.\ \cite{Goldberg:1986nk,Foot:2003iv}.  Fig.\
\ref{epsfig} shows that much larger values of $\epsilon$ are needed to
stop the ions in the earth than to detect them.

\subsection{Special case $m_\de = m_\pd$}

If for some reason ({\it e.g.,} a discrete symmetry) $m_\de = m_\pd$,
the matrix element for elastic scattering vanishes in the Born
approximation since the average charge density in the $\dH$ atom
vanishes.  This is an interesting situation since then inelastic
scattering can be the dominant effect for direct detection.  The
hyperfine splitting is given by
\be
	E_{\rm hf} = \sfrac23\, g_\de\, g_\pd\, \alpha'^4 {m_\de^2 m_\pd^2\over
	(m_\de+m_\pd)^3} \to \sfrac16 \alpha'^4 m_\dH
\ee
assuming gyromagnetic ratios $g_\de = g_\pd = 2$ and $m_\dH \cong
2m_\pd$.
The transitions from the spin singlet to triplet atomic DM states
are dominated by the spin-orbit coupling between the proton and the
constituents of $\dH$,
\bea
H_{\rm int} &=& {\tilde\epsilon e\over 4\pi\, m_p\,r^3}
 \vec L_p\cdot \vec\mu_\de +\  \{ \de \to \pd \}
\eea
where $\vec\mu_\de = g\vec\sigma_\de / m_\de$ is the dark magnetic
moment of $\de$ (hence $\tilde\epsilon\vec\mu_\de = 
\epsilon e\vec\sigma_\de / m_\de$ is its normal magnetic
moment) and $r$ is the distance between $p$ and $\de$. (Notice there
is no reduction by $1/2$ for Thomas precession since the electron 
rest frame is effectively inertial.) We neglect
the spin-spin couplings because these give rise to spin-dependent
interactions with the nucleus that are suppressed due to the lack of 
coherence.  

The squared matrix element, summed over the final spin states of the
triplet, and taking into account the equal contributions from
$\de$ and  $\pd$, is given by
\bea
	\sum_s|\langle \vec p\,',s|H_{\rm int}|\vec p,0\rangle|^2 
	\!\!&=&\!\!
 	{C^2\over 4 \pi^2}\!\sum_s\! \left|\langle 
	\vec p\,'|{\vec L_p\over m_p r^3}
	|\vec p\rangle\cdot \langle s | \vec\sigma_\de| 0\rangle
	\right|^2 \nonumber\\
	&=&  {C^2  \mu_{p}\over p\, q^2} {|\vec v\times\hat q|^2\over
	(1 + q^2 {a'_0}^2/4)^4}
\eea
where  $\vec p$, $\vec p\,'$ are the initial and final momenta of
the proton (in the rest frame of $\dH$), $\vec q = \vec p - \vec p\,'$
is the momentum transfer, $C = \epsilon e^2/ m_\de$,  and
$\mu_{p}/p = m_p m_\dH/[(m_p+m_\dH)p]$ is from normalizing the incoming plane wave
to unit current density \cite{LL}.  From this we obtain the inelastic
differential cross section for protons on dark atoms,
which can be rescaled to represent the nucleus-atom cross section
by including the factor of $Z^2$ to sum over the individual proton
contributions, letting $\mu_p\to \mu_N$, and inserting the Helm form factor $F_H^2(q)$ to account for the
nuclear structure:
\be
	{d\sigma_N \over d\Omega} \cong {(4\epsilon Z)^2
\alpha^2\over m_\dH^2}\, {\mu_N^2\over q^2}\,
	{p'\over p} |\vec v\times\hat q|^2 F_H^2
\label{us}
\ee
where now $p$, $p'$ stand for the initial and final momenta of $\dH$
(in the lab frame), $\vec q = \vec p-\vec p\,'$ as before, and we
used $m_\de = \sfrac12 m_\dH$.

Let us compare to the corresponding result (40, 42) of 
\cite{Kaplan:2009de} in the $m_\de = m_\pd$ limit:
\be
	{d\sigma_N \over d\Omega} \sim  {4\alpha\over \pi}
	Z^2\, {\mu_N^2\over m_p^2}\, {q^2\over f_{\rm eff}^4} F_H^2
\label{them}
\ee
By equating (\ref{us}) with (\ref{them}) using the preferred values
for fitting to the CoGeNT data, $A = 73$, $q = \sqrt{2 m_N E_R}
\sim 26$ MeV for recoil energy $E_R = 5$ keV, $v\sim 2\times
10^{-3}c$, and 
$f_{\rm eff}^{-2} \sim
10^{-30}$ cm$^2$ \cite{Kaplan:2011yj}, we find that 
\be
\epsilon  \sim 10^{-2}
\label{epseq}
\ee
to explain CoGeNT.
The splitting $E_{\rm hf} = 15$ keV and atomic mass $m_\dH = 6$ GeV imply
$\alpha' = 0.062$.
We confirm the estimate (\ref{epseq}) by comparing  $\sigma_p
\sim \pi[8 \epsilon\,\alpha\, \mu_p v/(q\,m_\dH)]^2$ ({\it c.f.}\ eq.\
(\ref{us})) to
the determination $\sigma_p\cong 10^{-38.3}$ cm$^2$ from 
fig.\ 12 of ref.\ \cite{Chang:2010yk}, which was the first to propose 
inelastic scattering (also with a mass difference of 15 keV) as an 
explanation for the CoGeNT observations. We note that dark ions will not penetrate
1 km of rock for such large $\epsilon$, considering fig.\ \ref{epsfig}.

\vspace{-0.25cm}
\section{Other Constraints}

{\bf Laboratory and supernova bounds.} 
Unlike models in which the $\gamma'$ has a mass, in ours no coupling
of the $\gamma'$ to visible sector matter is induced by the kinetic
mixing.  Therefore a variety of bounds that would pertain
to massive $\gamma'$s, from beam-dump experiments,
contributions to the anomalous
magnetic dipole moments of the electron and muon, 
and supernova emission of $\gamma'$,  do not apply here. Note that the dark matter is too
heavy to be in equilibrium in supernovae.  Accelerator constraints for
millicharged particles have a large open window for $\epsilon< 0.1$
and masses $\gsim 1$ GeV of interest here \cite{Jaeckel:2010ni}.

{\bf Exotic isotopes.}  Millicharged dark ions  with ionization
fraction  $f=10^{-4}$ would ostensibly have a flux of $2\times 10^{20}$/s on
the earth, and they would bind to normal nuclei unless they are unstable
against thermal fluctuations, requiring $\epsilon\lsim 10^{-3}$
\cite{Goldberg:1986nk,Holdom:1986eq}.  With $\epsilon\sim 10^{-2}$,
they would be stopped in $\sim 1$ m of the atmosphere ($10^{44}$ atoms) and produce
a relative abundance $10^{-7}$ of exotic isotopes over $10$ Gyr. In
the mass range  covering $m_\de = 3$ GeV, heavy isotope searches have
excluded abundances of $10^{-18.5}$ for deuterium from D$_2$0
\cite{Muller:1976qy} and $10^{-14}$ for helium \cite{klein}.  

However, there are a number of reasons why these limits would not
apply to our model. We evade the first one because $\de$ binds much
more strongly (400 eV) to oxygen than to deuterium (3 eV) for
$\epsilon =0.01$, making D[D$\de$]O highly unstable to decay into
D$_2$[O$\de$].  The second is ameliorated by realizing that He has a
lifetime of $\tau = 10^6$ y in the atmosphere 
\cite{Lu:2004pk}, reducing the estimated
abundance by a factor of $\tau/(10$ Gyr) to $10^{-11}$.  This does not
yet take into account the magnetosphere which very effectively shields
the earth from slow charged particles, including 3 GeV dark ions with 
$\epsilon\sim 10^{-2}$, whose gyroradius at the top of the atmosphere
is $\sim 0.01$ earth radii.  Solar x-rays are sufficiently energetic
to break up the He-$\de$ bound state (binding energy 5 eV) and allow
$\de$ to rebind much more strongly to N or O in the atmosphere.  Even
though ref.\ \cite{Chuzhoy:2008zy} conservatively limits 
$\epsilon < 0.005$ for 
supernovae to be able to efficiently expel $3$ GeV ions  
from the galaxy, which is
  marginally smaller than our
preferred value, the galactic $B$ field does prevent new ions from
entering the galaxy, and this could decrease the expected flux. 
Moreover our determination of $\epsilon$ could decrease by a factor of 
$10^{0.5}$ in light of CoGeNT's recent reanalysis of their
background events \cite{Collar}. Thus
it is far from clear that exotic isotope searches rule out our
$\epsilon\sim 10^{-2}$, $m_\de=m_\pd=3$ GeV model.

{\bf Big bang nucleosynthesis (BBN).} 
The current bound of no more than one additional neutrino 
\cite{Mangano:2011ar} in the plasma at BBN gives  weak constraints on
$\epsilon$ since the extra $\gamma'$ counts as only $8/7$ of a
neutrino.  Thus even if $\gamma'$ remains completely in equilibrium,
it only exceeds the 95\% c.l.\ bound by $0.14$ of an additional
neutrino species.  As expected, the constraints on $\epsilon$ are
quite weak from demanding decoupling of processes keeping $\gamma'$ in
equilibrium early enough to dilute this small fraction of a species.

{\bf Neutron stars.}
It has been pointed out that asymmetric bosonic dark matter must
interact extremely weakly with baryons in order to avoid the
destruction of neutron stars due to accumulation leading to 
gravitational collapse \cite{McDermott:2011jp,Kouvaris:2011fi,Guver:2012ba}.
This restricts the cross section to values below that needed for
direct detection, but it relies upon the absence of Fermi pressure
for bosonic DM.  Once the atoms become so dense that they
are overlapping, their constituents start to behave as a Fermi gas,
and thus avoid collapse until much higher densities, as long as the
overlap of atoms occurs before they are within their Schwarzschild
radius, $R_s$.  For $N$ dark atoms of mass $m_\dH$, $R_s = 2 N m_\dH/
M_p^2$ and the average separation between atoms is $\Delta r =
R_s/N^{1/3}$.  The criterion to avoid collapse is that $\Delta
r\ll a_0'$ at the bosonic Chandrasekhar limit $N\sim (M_p/m_\dH)^2$ \cite{McDermott:2011jp}.    We find that 
$\Delta r/a_0' \sim \alpha' m_\de/(m_\dH^{1/3} M_p^{2/3})\ll 1$, so
the system is no longer atomic, but a plasma of dark ions.  In this
case the much weaker fermionic Chandrasekhar limit $N \sim
M_p^3/(m_\de^{3/4} m_{\pd}^{9/4})$ \cite{guy} applies, and there is
no constraint from neutron stars.  ($\Delta r/a_0' \sim
\alpha'\sqrt{m_\de/m_\pd}$ is still self-consistently $< 1$ in this
case.)

{\bf Halo shape constraints.}
The most serious constraints on atomic dark matter self-interactions
come from their distortions of the shapes of elliptical DM cores
of galactic clusters.  To avoid relaxation to more spherical shapes,
ref.\ \cite{MiraldaEscude:2000qt} finds the constraint 
$\sigma/m_\dH < 0.02$ cm$^2$/g
on the cross section for $\dH$-$\dH$ collisions.  Ref.\ 
\cite{Kaplan:2009de}
argues that $\sigma = 4\pi (\kappa a_0')^2$ with $3\lsim \kappa\lsim
10$.  Together with the bound on $\sigma$, this would rule out our
inelastic model with $m_\dH = 6$ GeV.  However, assuming that the
elastic cross section for $\dH$-$\overline\dH$ scattering computed in 
\cite{froelich}
is the same as for $\dH$-$\dH$, we find that $\kappa \cong 0.16$
at the relevant energy $E \sim
(v/\alpha')^2\sim 10^{-3}$ in atomic units.  
The ellipticity constraint is then 
satisfied for the inelastic model, with $m_\dH\gsim 2$ GeV.  For the elastic model
with $\alpha' = m_\de/m_\pd = 0.1$, it implies $m_\dH \gsim 4$ GeV.
Bounds on $\sigma$ from the Bullet Cluster give weaker constraints
\cite{Randall:2007ph}.

{\bf Cosmic microwave background (CMB).} 
Bounds on $\epsilon$ were obtained on millicharged DM in ref.\
\cite{McDermott:2010pa} considering a variety of physical processes;
the most stringent limits were obtained from demanding that dark
matter has decoupled from the photon-baryon plasma before
recombination, under the assumption that the DM was fully ionized.  
These are
weakened in our model due to the screened electromagnetic interactions
of the atoms or the ionization fraction $f$ being small.
By extrapolating the results of fig.\ 1 of ref.\ \cite{Kaplan:2009de}
to  $m_\de = m_\pd = 3$ GeV we find that $f = 10^{-4}$ at $\alpha' =
0.062$.  This is well below the limit $f = \Omega_{\rm
ion}/\Omega_{\rm atom}
< 0.007/0.11$ from distortions of the cosmic microwave background
\cite{Dubovsky:2003yn}.  One also requires that the DM be out of
kinetic equilibrium with the baryon-photon plasma before
recombination.  For the atoms, the dominant interaction is
$\gamma \dH\to\gamma\dH$ through the Compton cross section
$\sigma_{\sss C} = {32\pi\epsilon^2 \alpha^2/3 m_\de^2}$.
We find the weak limit $\epsilon < 0.02$, which is marginally
consistent with (\ref{epseq}).
Thus we evade the strongest constraints
on $\epsilon$ in fig.\ 1 of \cite{McDermott:2010pa} in our inelastic
scattering model with $m_\de = m_\pd$.

In the case $m_\de < m_\pd$ where elastic scattering dominates 
in direct detection, we need not consider how much the bounds
of ref.\ \cite{McDermott:2010pa} are softened by having dark atoms
rather than ions.  In this case, the Xenon100 limit on $\epsilon$,
at least for the model $\alpha' = m_\de/m_\pd = 0.1$, is well below
the most stringent limit of \cite{McDermott:2010pa} even if the dark
matter was fully ionized.  This is shown in fig.\ \ref{epsfig}.

{\bf Virialization of dark matter.}
The next-strongest bounds in \cite{McDermott:2010pa} arise from heating
of the DM by scattering from baryons, which would interfere with DM
collapse and virialization in galaxy formation.  However in the
$m_\de=m_\pd$ case, the tree-level exchange of photons between $\dH$
and normal H atoms vanishes because of the perfect mutual screening of
$\de$ and $\pd$.  Thus there is no efficient way of
heating up the dark atoms; only the small ionized fraction  suffers
this fate, and it should have a negligible effect on galaxy
formation. For the $m_\de\ll m_\pd$ case, our Xenon100 bound on 
$\epsilon$ is more restrictive than the virialization bound.



\section{Conclusions}

We have reconsidered a minimal alternative for atomic dark
matter, namely that at the Lagrangian level its ionic constituents
couple only to one massless  $\gamma'$ gauge boson, responsible for
their binding, but due to kinetic mixing of the
$\gamma'$ with the photon, they acquire small charges $\pm\epsilon
e$ giving the dark atoms a weak coupling to the ordinary photon.  We have shown
that $\epsilon$ is sufficiently unconstrained so that the first
evidence of this interaction might come through direct detection of
the dark atom. 

There are two interesting regions of parameter space for the model
with respect to direct detection.
If  the dark ``electron'' is much lighter than the dark ``proton,''
$m_\de \ll m_\pd$, then dark atoms scatter primarily elastically on
nuclei, and the cross section can be close to the limits from direct
detection over a wide range of masses $m_\dH$.  However if $m_\de =
m_\pd$, elastic scattering is suppressed and inelastic hyperfine
transitions dominate.  We find that if the $\gamma'$ gauge coupling is
$\alpha' = 0.06$ and $m_\de = m_\pd\cong 3$ GeV, the hyperfine splitting is
$\sim 15$ keV, and $\epsilon \cong 10^{-2}$ gives the right cross
section for explaining candidate DM events reported by CoGeNT.

This model has further testable implications. The detailed effects of
atomic DM on the CMB, depending upon exactly how and when it
recombines, have yet to be studied. An important unexplored aspect  is
the fraction of atoms that combine into molecular $\dH_2$.  This could
result in multiple signals for direct detection, where dark atoms or
molecules  interact with nuclei, producing recoils at energies
corresponding to several different DM masses.  Searches for
exotic isotopes consisting of  bound states of $\de$ and normal atoms
appear to be tantalizingly close to ruling out the $\epsilon=0.01$
model, but more work must be done to quantify the predicted
abundances.    The framework
appears to be consistent with  strong DM halo ellipticity constraints,
but an independent determination of the $\dH$-$\dH$ scattering cross
section at the relevant energies should be done to confirm this.  The
origin of the DM asymmetry  \cite{Kaplan:2011yj} should also be
addressed. We hope to return to these issues in the near future.

\noindent{\bf Acknowledgements.}  
We thank S.\ Davidson, R.\ Foot, D.E.\ Kaplan, G.\ Krnjaic,
A.\ Kurkela, 
S.\ McDermott, 
G.\ Moore, M.\ Pospelov, P.\ Scott,  N.\ Toro, and K.\ Zurek
for helpful interactions.  We are supported by NSERC.


\begin{thebibliography}{10}

\bibitem{Foot:2003iv} 
  R.~Foot,
  Phys.\ Rev.\ D {\bf 69}, 036001 (2004).

\bibitem{Huh:2007zw} 
  J.~-H.~Huh, J.~E.~Kim, J.~-C.~Park and S.~C.~Park,
  Phys.\ Rev.\ D {\bf 77}, 123503 (2008)
  [arXiv:0711.3528 [astro-ph]].

\bibitem{Pospelov:2007mp} 
  M.~Pospelov, A.~Ritz and M.~B.~Voloshin,
  Phys.\ Lett.\ B {\bf 662}, 53 (2008)
  [arXiv:0711.4866 [hep-ph]].



\bibitem{ArkaniHamed:2008qn} 
  N.~Arkani-Hamed {\it et al.}, D.~P.~Finkbeiner, T.~R.~Slatyer and N.~Weiner,
  Phys.\ Rev.\ D {\bf 79}, 015014 (2009)




\bibitem{Alves:2009nf} 
  D.\ S.\ M.\ Alves, S.\ R.\ Behbahani, P.\ Schuster and J.\ G.\ Wacker,
  Phys.\ Lett.\ B {\bf 692}, 323 (2010)
  [arXiv:0903.3945];
  JHEP {\bf 1006}, 113 (2010)
  [arXiv:1003.4729].

\bibitem{Chen:2009ab} 
  F.~Chen, J.~M.~Cline and A.~R.~Frey,
  Phys.\ Rev.\ D {\bf 80}, 083516 (2009)
  [arXiv:0907.4746 [hep-ph]].

\bibitem{Feldman:2007wj} 
  D.~Feldman, Z.~Liu and P.~Nath,
  Phys.\ Rev.\ D {\bf 75}, 115001 (2007)
  [hep-ph/0702123].



\bibitem{Goldberg:1986nk} 
  H.~Goldberg, L.~J.~Hall,
  Phys.\ Lett.\ B {\bf 174}, 151 (1986).

\bibitem{Feng:2009mn} 
  J.~L.~Feng, M.~Kaplinghat, H.~Tu and H.~-B.~Yu,
  JCAP {\bf 0907}, 004 (2009)
  [arXiv:0905.3039 [hep-ph]].


\bibitem{Kaplan:2009de} 
  D.~E.~Kaplan, G.~Z.~Krnjaic, K.~R.~Rehermann and C.~M.~Wells,
  JCAP {\bf 1005}, 021 (2010)
  [arXiv:0909.0753].


\bibitem{Kaplan:2011yj} 
  D.~E.~Kaplan, G.~Z.~Krnjaic, K.~R.~Rehermann and C.~M.~Wells,
  JCAP {\bf 1110}, 011 (2011)
  [arXiv:1105.2073].

\bibitem{Kaplan:2009ag} 
  D.~E.~Kaplan, M.~A.~Luty and K.~M.~Zurek,
  Phys.\ Rev.\ D {\bf 79}, 115016 (2009)
  [arXiv:0901.4117 [hep-ph]].

\bibitem{Ackerman:2008gi} 
  L.~Ackerman, M.~R.~Buckley, S.~M.~Carroll and M.~Kamionkowski,
  Phys.\ Rev.\ D {\bf 79}, 023519 (2009)

\bibitem{Aalseth:2010vx} 
  C.~E.~Aalseth {\it et al.}  [CoGeNT Collaboration],
  Phys.\ Rev.\ Lett.\  {\bf 106}, 131301 (2011)
  [arXiv:1002.4703].

\bibitem{Angle:2011th} 
  J.~Angle {\it et al.}  [XENON10 Collaboration],
  Phys.\ Rev.\ Lett.\  {\bf 107}, 051301 (2011)
  [arXiv:1104.3088].

\bibitem{Holdom:1985ag} 
  B.~Holdom,
  Phys.\ Lett.\ B {\bf 166}, 196 (1986).

\bibitem{Burrage:2009yz} 
  C.~Burrage, J.~Jaeckel, J.~Redondo and A.~Ringwald,
  JCAP {\bf 0911}, 002 (2009)
  [arXiv:0909.0649 [astro-ph.CO]].


\bibitem{Aprile:2011hi} 
  E.~Aprile {\it et al.}  [XENON100 Collaboration],
  Phys.\ Rev.\ Lett.\  {\bf 107}, 131302 (2011)
  [arXiv:1104.2549].

\bibitem{McDermott:2010pa} 
  S.~D.~McDermott, H.~-B.~Yu and K.~M.~Zurek,
  Phys.\ Rev.\ D {\bf 83}, 063509 (2011)
  [arXiv:1011.2907 [hep-ph]].

\bibitem{Chuzhoy:2008zy} 
  L.~Chuzhoy and E.~W.~Kolb,
  JCAP {\bf 0907}, 014 (2009)
  [arXiv:0809.0436 [astro-ph]].

\bibitem{Foot:2010yz} 
  R.~Foot,
  Phys.\ Lett.\ B {\bf 699}, 230 (2011)
  [arXiv:1011.5078].



\bibitem{LL}
L.D.\ Landau and E.M.\ Lifshitz, {\it 
 vol.\ 3, 
Quantum Mechanics, Nonrelativistic
Theory}, Pergamon Press, 1977

\bibitem{Chang:2010yk} 
  S.~Chang, J.~Liu, A.~Pierce, N.~Weiner and I.~Yavin,
  JCAP {\bf 1008}, 018 (2010)
  [arXiv:1004.0697 [hep-ph]].

\bibitem{Jaeckel:2010ni} 
  J.~Jaeckel and A.~Ringwald,
  Ann.\ Rev.\ Nucl.\ Part.\ Sci.\  {\bf 60}, 405 (2010)
  [arXiv:1002.0329 [hep-ph]].

\bibitem{Holdom:1986eq} 
  B.~Holdom,
  Phys.\ Lett.\ B {\bf 178}, 65 (1986).

\bibitem{Muller:1976qy} 
  R.~A.~Muller, L.~W.~Alvarez, W.~R.~Holley and E.~J.~Stephenson,
  Science {\bf 196}, 521 (1977).

\bibitem{klein} J.\ Klein, R.\ Middleton and W.E.\ Stephens,
in ANL Symposium on Accelerator Mass Spectroscopy
(1981)

\bibitem{Lu:2004pk} 
  Z.~T.~Lu {\it et al.,} 
  Nucl.\ Phys.\ A {\bf 754}, 361 (2005)




\bibitem{Collar}
J.\ Collar, TAUP 2011 


\bibitem{Mangano:2011ar} 
  G.~Mangano and P.~D.~Serpico,
  Phys.\ Lett.\ B {\bf 701}, 296 (2011)
  [arXiv:1103.1261 [astro-ph.CO]].

\bibitem{McDermott:2011jp} 
  S.~D.~McDermott, H.~-B.~Yu and K.~M.~Zurek,
  arXiv: 1103.5472 [hep-ph].

\bibitem{Kouvaris:2011fi} 
  C.~Kouvaris and P.~Tinyakov,
  Phys.\ Rev.\ Lett.\  {\bf 107}, 091301 (2011)
  [arXiv:1104.0382 [astro-ph.CO]].


\bibitem{Guver:2012ba} 
  T.~Guver, A.~E.~Erkoca, M.~H.~Reno and I.~Sarcevic,
  arXiv:1201.2400 [hep-ph].

\bibitem{guy}
G.D.\ Moore, private communication

\bibitem{MiraldaEscude:2000qt} 
  J.~Miralda-Escude,
  Ap.~J.~{\bf 564} (2002) 60
  [astro-ph/ 0002050].


\bibitem{froelich}
P.\ Froelich {\it et al.,} Phys.\ Rev.\ Lett.\ {\bf 84} 4577 (2000)

\bibitem{Randall:2007ph} 
  S.~W.~Randall {\it et al.,}
  Astrophys.\ J.\  {\bf 679}, 1173 (2008)
  [arXiv:0704.0261 [astro-ph]].


\bibitem{Dubovsky:2003yn} 
  S.~L.~Dubovsky, D.~S.~Gorbunov and G.~I.~Rubtsov,
  JETP Lett.\  {\bf 79}, 1 (2004)
  [hep-ph/0311189].




\end{thebibliography}
\end{document}